\documentclass[10pt,conference]{IEEEtran}
\IEEEoverridecommandlockouts

\usepackage[T1]{fontenc}
\usepackage[utf8]{inputenc}
\usepackage{float}
\usepackage{multirow}
\usepackage{amsthm}
\usepackage{graphicx}
\usepackage{changepage}
\usepackage{url}
\usepackage{amssymb}
\PassOptionsToPackage{normalem}{ulem}
\usepackage{ulem}
\usepackage{array}
\usepackage{booktabs}
\usepackage[table,x11names]{xcolor}
\usepackage{amsmath}
\usepackage{stackrel}
\usepackage{babel}
\usepackage{amsmath}
\usepackage{amsthm}
\usepackage{pgfplots}
\usepackage{pgf-pie}
\usepackage[table]{xcolor}
\usepackage{enumitem}
\usepackage{balance}
\usepackage{hyperref}
\usepackage{xcolor}

\usepackage{pgfplots}
\usepackage{pgfplotstable}
\pgfplotsset{compat=1.7}
\usepackage{tikz}

\hypersetup{
  colorlinks   = True, 
  urlcolor     = blue, 
  linkcolor    = blue, 
  citecolor    = blue 
}

\begin{document}
\title{Architecture Decisions in AI-based Systems Development: An Empirical Study}

\author{
    \IEEEauthorblockN{Beiqi Zhang$^{1}$, Tianyang Liu$^2$, Peng Liang$^{1*}$\thanks{\indent This work is funded by the NSFC with Grant No. 62172311.}, Chong Wang$^{1*}$, Mojtaba Shahin$^2$, Jiaxin Yu$^{1}$}
    \IEEEauthorblockA{$^1$ School of Computer Science, Wuhan University, Wuhan, China}
    \IEEEauthorblockA{$^2$ Department of Computer Science and Engineering, University of California San Diego, La Jolla, USA}
    \IEEEauthorblockA{$^3$ School of Computing Technologies, RMIT University, Melbourne, Australia}
    \IEEEauthorblockA{\{\href{mailto:liangp@whu.edu.cn}{liangp}, \href{mailto:cwang@whu.edu.cn}{cwang}\}@whu.edu.cn}
}







%
\maketitle       

\begin{abstract}
Artificial Intelligence (AI) technologies have been developed rapidly, and AI-based systems have been widely used in various application domains with opportunities and challenges. However, little is known about the architecture decisions made in AI-based systems development, which has a substantial impact on the success and sustainability of these systems. To this end, we conducted an empirical study by collecting and analyzing the data from Stack Overflow (SO) and GitHub. More specifically, we searched on SO with six sets of keywords and explored 32 AI-based projects on GitHub, and finally we collected 174 posts and 128 GitHub issues related to architecture decisions. The results show that in AI-based systems development (1) architecture decisions are expressed in six linguistic patterns, among which \textit{Solution Proposal} and \textit{Information Giving} are most frequently used, (2) \textit{Technology Decision}, \textit{Component Decision}, and \textit{Data Decision} are the main types of architecture decisions made, (3) \textit{Game} is the most common application domain among the eighteen application domains identified, (4) the dominant quality attribute considered in architecture decision-making is \textit{Performance}, and (5) the main limitations and challenges encountered by practitioners in making architecture decisions are \textit{Design Issues} and \textit{Data Issues}. Our results suggest that the limitations and challenges when making architecture decisions in AI-based systems development are highly specific to the characteristics of AI-based systems and are mainly of technical nature, which need to be properly confronted.
\end{abstract}

\begin{IEEEkeywords}
Architecture Decision, AI-based Systems Development, Stack Overflow, GitHub, Empirical Study
\end{IEEEkeywords}

\section{Introduction}
\label{sec:introduction}

Artificial Intelligence (AI) is the science and engineering of making intelligent machines complete tasks using human intelligence \cite{mccarthy2004artificial}. AI-based systems are software systems with functionalities enabled by at least one AI component (e.g., for image recognition and autonomous driving) \cite{Fernandez2021}. In the last decade, with the support of massively parallel computer systems, huge datasets, and better algorithms, AI has brought a number of important applications to near-human levels of performance, which simultaneously enabled the advances in AI technologies \cite{anthes2017artificial}. AI is growing at a rapid pace and will be the most transformative and potential field of the next decade \cite{anthes2017artificial}. AI technologies have been extensively used in education, medical treatment, elderly care, environmental protection, urban operation, judicial services, etc. As a result, the application of AI-based systems is now becoming an integrated part of our daily life.

Software architecture is the blueprint for software development and plays a critical role in managing the complex interactions between stakeholders in software-intensive systems to balance various constraints (e.g., system quality requirements) \cite{Bass2012Software}. Designing the architecture of a software system can be viewed as a decision-making process, and software architecture is fundamentally a combination of design decisions on which decisions are made in response to the essential requirements \cite{Jansen2005Software}. Making informed architecture decisions during the development process leads to better control over the design, development, and evolution of large and dynamic software systems \cite{Shahin2009Architeccture}, such as microservices systems~\cite{waseem2022decision}. 
Although software architecture and architecture decisions have been extensively studied in the last two decades, little research (e.g., \cite{franch2022architectural}) has been conducted on architecture decisions in AI-based systems development. Franch \textit{et al.}~\cite{franch2022architectural} have made a literature survey on this area, and they highlighted the need to further investigate the current state of research and practice on architecture decisions in AI-based systems development. AI-based systems are different from traditional software-intensive systems as they are more complex, include new stakeholders (e.g., Machine Learning (ML) developers, ethics experts) with new concerns (e.g., ethics, ML model accuracy), and have new components (e.g., components with ML capabilities) \cite{muccini2021software,lewis2021characterizing, humbatova2020taxonomy}. All these make architecting AI-based systems more challenging \cite{muccini2021software, carleton2021architecting}, and necessitate making informed architecture decisions to address new challenges associated with AI-based systems.

To this end, we conducted a qualitative empirical study that collected data from Stack Overflow (SO) and GitHub, two popular developer communities, as the data sources to get the opinions of practitioners about architecture decisions in AI-based systems development. The reasons are that SO is the most popular website where software practitioners often look for solutions \cite{stack2022survey}, leading to a wide collection of development-related information discussed at SO, and GitHub has hosted a large number of AI-based projects where issue tracking systems provide a vehicle for developers to discuss issues and solutions. Considering this, we set SO and GitHub as our data sources and collected data about architecture decisions made in AI-based systems development. We then used both a predefined classification and the Constant Comparison method to analyze the data items extracted.

\textbf{Our findings show that} (1) \textit{Solution Proposal} and \textit{Information Giving} are the most frequently used linguistic patterns, (2) \textit{Technology Decision}, \textit{Component Decision}, and \textit{Data Decision} are the main types of architecture decisions, (3) \textit{Game} is the most common application domain identified, (4) \textit{Performance} is the dominant quality attribute considered, and (5) \textit{Design Issues} and \textit{Data Issues} are the main limitations and challenges when making architecture decisions in AI-based systems development.

\textbf{The contributions of this work}: (1) we identified the linguistic patterns practitioners used to express architecture decisions in AI-based systems development; (2) we explored various aspects of architecture decisions in AI-based systems development, including the types of architecture decisions made, application domains, and the quality attributes considered in architecture decision-making; (3) we analyzed and categorized the challenges and limitations practitioners encountered when making architecture decisions in AI-based systems development, which are potential directions to be explored for researchers; (4) we provided suggestions for practitioners to make informed architecture decisions in AI-based systems development; and (5) we released a dataset on architecture decisions made in AI-based systems~\cite{replpack}.

The rest of this paper is structured as follows: Section \ref{sec:mapping} presents the research questions and the research design of this study. Section \ref{sec:results} provides the study results, which are further discussed in Section \ref{sec:discussion}. The potential threats to validity are clarified in Section \ref{sec:threats}. Section \ref{sec:relatedWork} surveys the related work and Section \ref{sec:conclusions} concludes this work with future directions.

\section{Methodology}
\label{sec:mapping}
\begin{figure*}[htbp]
	\centering
	\includegraphics[width=1.0\linewidth]{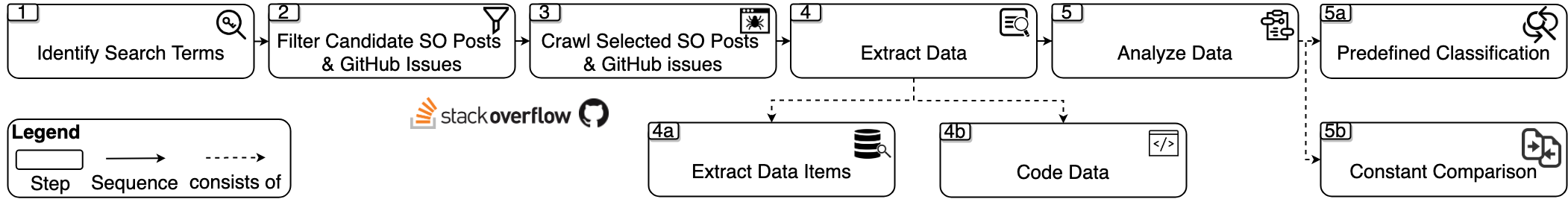}
	\caption{Overview of the research process}
	\label{fig:Overview of research process}
\end{figure*}

The goal of this study, described by the Goal-Question-Metric approach \cite{Basili1994TheGQ}, is to \textbf{analyze} \textit{architecture decisions in AI-based systems development} \textbf{for the purpose of} \textit{characterization} \textbf{with respect to} \textit{decision description and categorization, application domains covered, quality attributes considered, and the limitations and challenges when making architecture decisions} \textbf{from the point of view of} \textit{practitioners} \textbf{in the context of} \textit{architecture related SO posts and GitHub issues of AI-based systems}. We conducted this empirical study by following the guidelines proposed in \cite{Easterbrook2008}. The Research Questions (RQs), their rationale, and the research process (see Figure \ref{fig:Overview of research process}) are explained in the subsections below.

\subsection{Research Questions} \label{Research Questions}
\label{RQs}
\noindent \textbf{RQ1: How do developers express architecture decisions made in AI-based systems development?}\\
\noindent \emph{Rationale:} In architecture decision-making of AI-based systems, developers generally use natural language to describe their ideas and communicate with each other. This RQ aims to provide the linguistic patterns used by practitioners to express architecture decisions in AI-based systems development, and the answer to this RQ can help developers identify architecture decision-related content in AI-based systems development.\\
\noindent \textbf{RQ2: What types of architecture decisions are made in AI-based systems development?}\\
\noindent \emph{Rationale:} Developers make a variety of architecture decisions when developing AI-based systems. However, little is known about what types of architecture decisions are made. The aim of this RQ is to provide a categorization of architecture decisions made in AI-based systems development.\\
\noindent \textbf{RQ3: What are the domains of AI-based systems in which architecture decisions are made?}\\
\noindent \emph{Rationale:} AI technologies have been extensively used to develop software systems in various application domains. The aim of this RQ is to get an overview of the application domains of AI-based systems in which architecture decisions are made, and the answer to this RQ can provide suggestions for making architecture decisions in specific domains. \\
\noindent \textbf{RQ4: What quality attributes are considered when developers make architecture decisions in AI-based systems development?}\\
\noindent \emph{Rationale:} Compared with traditional software systems, AI-based systems have strong data dependencies, high coupling, and uncertainty. Therefore, the quality attributes that need to be considered for AI-based systems are different from traditional software systems \cite{ozkaya2020what}. This RQ aims to provide an understanding of the quality attributes currently considered when making decisions in AI-based systems development.\\
\noindent \textbf{RQ5: What are the limitations and challenges of making architecture decisions in AI-based systems development?}\\
\noindent \emph{Rationale:} Although AI technologies have been widely used in the software industry, and practitioners are becoming increasingly skilled in developing AI-based systems, there are still restrictions and problems during architecture decision-making of AI-based systems. This RQ aims to collect and identify the limitations and challenges practitioners may experience when making architecture decisions for software systems with AI-related components. The answers of this RQ can provide researchers the potential directions to be further explored.\\

\subsection{Data Collection}
\label{sec:subject_projects} 
As one of the online software development communities, Stack Overflow (SO) has been the most popular and widely used question and answer (Q\&A) platform for developers to ask and answer questions. According to Stack Exchange, as of March 2022, SO has over 18 million registered users, and has received over 23 million questions and 33 million answers, which provides a wealth of development related data. GitHub is a provider of hosting for software development and version control using Git, from which the data can be complementary to the data from SO. Therefore, we decided to use SO and GitHub as the data sources to answer our RQs. More specifically, we used SO posts and GitHub issues to obtain practitioners' opinions on architecture decisions in AI-based systems development.

\subsubsection{Stack Overflow Data Collection}
To locate the SO posts related to architecture decisions in AI-based systems development, we started by conducting a pilot search process to formulate the search terms, and the search terms were applied to all parts of the posts, including the title, the body of the question, and the answers. First, we defined several terms to contain posts referring to AI-based systems, including ``\textit{artificial intelligence}'', ``\textit{AI}'', ``\textit{machine learning}'', and ``\textit{deep learning}''. We did not use search terms of specific AI techniques (e.g., ``\textit{reinforcement learning}'') or application areas (e.g., ``\textit{expert system}'') because these terms may lead to the bias of the search results to these techniques and application areas. Second, as SO supports the use of wildcard (*) searches to broaden the search results, we defined ``\textit{architect*}'' as the initial search term related to architecture decisions. Since this study is about architecture decisions in AI-based systems development, we formulated the search terms by combining the two subtopics as (``\textit{artificial intelligence}'' OR ``\textit{AI}'' OR ``\textit{machine learning}'' OR ``\textit{deep learning}'') AND (``\textit{architect*}''). Considering that we have already determined the search terms for AI-based systems, we only needed to confirm whether there are other variations of terms for architecture decisions in addition to the initial search term ``\textit{architect*}''.

The effectiveness of a keyword-based mining strategy highly depends on the appropriateness of the set of keywords selected. Hence, we followed the systematic approach proposed by Bosu \textit{et al.}~\cite{Bosu2014Identifying} to identify the keywords required for our study throughout the keyword-based search process on SO, and to see the existence of other terms referring to architecture decisions besides ``\textit{architect*}''. This includes the steps as follows:

\begin{enumerate}
    \item We developed an initial keyword set including (``\textit{artificial intelligence}'' OR ``\textit{AI}'' OR ``\textit{machine learning}'' OR ``\textit{deep learning}'') AND (``\textit{architect*}'').
    \item We built a corpus by using the initial set of keywords to search on SO and identified posts that contain at least one set of keywords (e.g., ``\textit{AI architect*}'').
    \item We converted all words in the posts identified in Step 1 to lowercase and removed all the numbers and punctuations. 
    \item We deleted stopwords in the corpus by applying the NLTK library~\cite{NLTK}.
    \item We created a list of tokens in the corpus.
    \item We used the Port stemming algorithm \cite{Willett2006The} to get each token's stem (e.g., \textit{works}, \textit{worked}, \textit{working} are all turned into \textit{work}).
    \item We created a Document-Term matrix \cite{Tan2005Introduction} from the corpus. 
    \item We looked for additional terms that regularly occur in conjunction with each of our initial keywords. If the probability of co-occurrence is greater than 0.05, the term can be considered for inclusion in the keyword set.
\end{enumerate}

After following the preceding eight steps, we did not discover any additional keywords whose co-occurring probability with our initial keyword sets was higher than 0.05. This boosted our confidence that our original keyword set is sufficient to search related posts on SO. Considering that practitioners often use the abbreviations ``\textit{ML}'' for ``\textit{machine learning}'' and ``\textit{DL}'' for``\textit{deep learning}'', we conducted a pilot search on SO using (``\textit{ML}'' OR ``\textit{DL}'') AND (``\textit{architect*}''). The number of posts retrieved by this search string was not very large, and it was easy to tell if the abbreviations in these posts refer to AI-based systems development. Therefore to make the keyword set more comprehensive, we decided to add ``\textit{ML}'' and ``\textit{DL}'' to our search terms in order to expand our search scope. The search terms that were utilized and the number of posts that were obtained from SO are reported in Table \ref{Search terms on Stack Overflow}. We got a total of 3,022 posts in this step. It is possible that a post contains multiple sets of keywords, which means that there could be duplicate items in the set of URLs of these posts. After removing the duplicated posts, we finally got 2,670 posts with unique URLs.

\begin{table}[htbp]
\caption{Search terms used in Stack Overflow}
\label{Search terms on Stack Overflow}
\begin{tabular}{m{0.8cm}<{\centering}m{4.3cm}m{2.5cm}<{\centering}}
\toprule
\textbf{\#}                                       & \textbf{Search Term}                                         & \textbf{Number of Retrieved Posts}  \\ \midrule
\textbf{ST1}                                      & ``\textit{AI architect*}''                                   & 413  \\
\textbf{ST2}                                      & ``\textit{artificial intelligence architect*}''              & 75   \\
\textbf{ST3}                                      & ``\textit{deep learning architect*}''                        & 770  \\
\textbf{ST4}                                      & ``\textit{machine learning architect*}''                     & 854  \\ 
\textbf{ST5}                                      & ``\textit{DL architect*}''                                   & 386  \\
\textbf{ST6}                                      & ``\textit{ML architect*}''                                   & 524  \\ \hline
\specialrule{0em}{1pt}{1pt} \textbf{Total}        &                                                              & 3022 \\ \bottomrule
\end{tabular}
\end{table}

\subsubsection{GitHub Data Collection}
First, we chose two existing datasets of AI-based systems from GitHub collected by Humbatova \textit{et al.}~\cite{humbatova2020taxonomy} and Gonzalez \textit{et al.}~\cite{gonzalez2020ML} as part of the source of AI-based systems in our study. Moreover, consulted with the literature \cite{humbatova2020taxonomy, gonzalez2020ML, nguyen2019machine}, we developed the keywords related to popular ML/DL frameworks in Python, C++, Java, and JavaScript, including “\textit{tensorflow}”, “\textit{keras}”, “\textit{torch}”, “\textit{sklearn}”, “\textit{caffe}”, “\textit{cntk}”, “\textit{mlpack}”, “\textit{deeplearning4j}”, “\textit{mllib}”, and “\textit{brain.js}”. It should be noted that the keywords were searched on the entire repository (i.e., the name, description, topics, and README file) using GitHub search API to identify AI-based projects. A repository might use different languages. Hence, we used a specific language name argument in our search, to restrict it to specific language source files (e.g., Java). We then merged these identified AI-based projects with the two existing datasets to form the final source of AI-based systems in this study. After this step, we got a number of 5,750 AI-based projects. We ranked these AI-based projects based on their number of issues and selected the top 32 AI-based projects. Note that we excluded the popular ML/DL frameworks, like PyTorch, because they are infrastructures mainly used to develop AI-based applications. The reason we chose the top 32 AI-based projects is that we planned to collect over 100 GitHub issues related to architecture decisions, which could be roughly satisfied by including the top 32 AI-based projects.

After identifying the 32 AI-based projects for collecting issue data, we used the keywords which are the same as the architecture search terms (i.e., ``\textit{architect*}'') used in Section~\ref{sec:subject_projects} \textit{1)}, ``\textit{architect}'', ``\textit{architecting}'', ``\textit{architected}'', ``\textit{architecture}'', and `` \textit{architectural}'' to search on the title and body of the closed issues of these 32 AI-based systems, which retrieved 2,341 closed issues containing the keywords. Table \ref{Projects in GitHub} shows the selected AI-based projects and their issues.


\begin{table}[htbp]
\caption{AI-based projects and their related architecture issues}
\label{Projects in GitHub}
\begin{tabular}{m{0.6cm}<{\centering}m{2.9cm}<{\centering}m{2cm}<{\centering}m{1.75cm}<{\centering}}
\toprule
\textbf{\#}     & \textbf{GitHub Project}   & \textbf{Number of Retrieved Issues}   & \textbf{Number of Related Issues}         \\ \midrule
\textbf{GP1}    & OpenCV                    & 251                               	& 4                                         \\
\textbf{GP2}    & Transformers              & 227	                                & 9                                         \\
\textbf{GP3}    & Deeplearning4J            & 150                               	& 10                                        \\
\textbf{GP4}    & spaCy                     & 576	                                & 6                                         \\
\textbf{GP5}    & Fairseq                   & 214	                                & 10                                        \\
\textbf{GP6}    & Kubeflow pipelines        & 21	                                & 7                                         \\
\textbf{GP7}    & AllenNLP                  & 63                                 	& 4                                         \\
\textbf{GP8}    & Distributed               & 26	                                & 2                                         \\
\textbf{GP9}    & MLflow                    & 21	                                & 9                                         \\
\textbf{GP10}   & DIGITS                    & 28	                                & 2                                         \\
\textbf{GP11}   & Triton Inference Server   & 64	                                & 6                                         \\
\textbf{GP12}   & Weights and Biases        & 26	                                & 2                                         \\
\textbf{GP13}   & Pyro                      & 12                                	& 1                                         \\
\textbf{GP14}   & Sagemaker Python SDK      & 6                                 	& 1                                         \\
\textbf{GP15}   & XLA                       & 35                                	& 6                                         \\
\textbf{GP16}   & TensorLayer               & 7                                 	& 2                                         \\
\textbf{GP17}   & MNE-Python                & 13	                                & 3                                         \\
\textbf{GP18}   & Orange                    & 9  	                                & 5                                         \\
\textbf{GP19}   & CNTK                      & 60	                                & 2                                         \\
\textbf{GP20}   & AirSim                    & 33	                                & 7                                         \\
\textbf{GP21}   & Shogun                    & 24                                	& 3                                         \\
\textbf{GP22}   & TensorFlow.js             & 83                                	& 3                                         \\
\textbf{GP23}   & NLTK                      & 9	                                    & 3                                         \\
\textbf{GP24}   & ONNX                      & 22                                	& 4                                         \\
\textbf{GP25}   & Face Recognition          & 7	                                    & 2                                         \\
\textbf{GP26}   & DVC                       & 23	                                & 1                                         \\
\textbf{GP27}   & DeepChem                  & 29                                	& 2                                         \\
\textbf{GP28}   & Gensim                    & 13                                	& 3                                         \\
\textbf{GP29}   & Tesseract                 & 56                                	& 1                                         \\
\textbf{GP30}   & cuML                      & 209                               	& 1                                         \\
\textbf{GP31}   & Bayeslite                 & 2	                                    & 2                                         \\
\textbf{GP32}   & OpenPAI                   & 20	                                & 5                                         \\ \hline    
\specialrule{0em}{1pt}{1pt} \textbf{Total}  & & 2341                                & 128                                       \\ \bottomrule
\end{tabular}
\end{table}

\subsection{Data Filtering}
To manually label SO posts and GitHub issues related to architecture decisions in AI-based systems development, we conducted a pilot data labelling by following a set of inclusion and exclusion criteria: (I1) If a post or issue is related to both architecture decisions and AI-based systems, we include it. (E1) If the focus of the question and answers in a post or an issue is only related to AI-based systems, but not related to architecture decisions, we exclude it. (E2) If the focus of the question and answers in a post or an issue is only related to architecture decisions, but not related to AI-based systems, we exclude it. Specifically, the pilot data labelling process is composed of the following steps: (1) The first author randomly selected 10 posts from the search results of each set of SO search terms (60 posts in total) and 10 issues from all the issues containing the keywords retrieved from the selected GitHub projects. (2) The first and third authors labelled independently whether the posts and issues should be included. (3) Data labelled by the two authors were compared, and the level of agreement for both SO posts and GitHub issues between the two authors were calculated using the Cohen's Kappa coefficient \cite{Jacob1960Coefficient}. (4) For any results that the two authors disagreed with, they were discussed by the two authors till an agreement was reached. The Cohen's Kappa coefficient before the discussion and resolution of the disagreements for labelling SO posts was 0.727 and 0.737 for GitHub issues, both higher than 0.7, and after the discussion, the two authors did not have any disagreements anymore, thus indicating a high degree of consistency between the two authors.

After the pilot data labelling, the first author checked all the posts retrieved from SO and all the issues retrieved from GitHub projects adhering to the inclusion and exclusion criteria used in the pilot data labelling. As it was easy to distinguish whether the posts and issues match the criteria, the first author completed this step alone. Only a few posts and issues that the first author was unsure were discussed with the third author. After manually filtering all the candidates, we got a total of 190 SO posts and 128 GitHub issues related to architecture decisions in AI-based systems development. As SO posts with the same content may have different URLs, we ended up with 174 posts after excluding such posts.

\subsection{Data Extraction and Analysis} 
\subsubsection{Extract Data}
The first author randomly selected 10 SO posts and 10 GitHub issues to conduct a pilot data extraction with the second author. The first two authors extracted the data items listed in Table \ref{Data Items and RQ} independently. If any disagreements were found, the third author was involved to discuss with the two authors and come to an agreement. After the pilot data extraction, the first two authors extracted the data items from the filtered posts and issues independently. During this process, they discussed the uncertain parts with the third author to ensure the accuracy of the extracted data. Finally, the first author compared and rechecked the extracted data by the two authors from all the collected posts and issues to further increase the correctness of the data.

\begin{table}[htbp]
\caption{Data items extracted and their corresponding RQs}
\label{Data Items and RQ}
\begin{tabular}{m{0.25cm}<{\centering}m{1.3cm}m{5cm}m{0.5cm}<{\centering}}
\hline
\textbf{\#} & \textbf{Data Item}             & \textbf{Description}                                                                                      & \textbf{RQ}   \\ \hline
D1          & Decision description           & \textit{Description of architecture decisions in AI-based systems development by practitioners based on their understanding}  
                                                                                                                                                                 & RQ1   \\ \hline
D2          & Decision type                  & \textit{A categorization of architecture decisions made in AI-based systems development}                          & RQ2   \\ \hline
D3          & Application domain             & \textit{Application domains of AI-based systems development in which architecture decisions are made}             & RQ3   \\ \hline
D4          & Quality attribute considered   & \textit{Considered quality attributes when practitioners making architecture decisions in AI-based systems development}      
                                                                                                                                                                 & RQ4   \\ \hline
D5          & Limitation and challenge       & \textit{The restrictions and difficulties when making architecture decisions in AI-based systems development}     & RQ5   \\ \hline
\end{tabular}
\end{table}

\subsubsection{Analyze Data}
We used a predefined classification and the Constant Comparison method to analyze the extracted data. Linguistic patterns are ``\textit{grammatical rules that allow their users to speak properly in a common language}''~\cite{da2017linguistic}. Sorbo \textit{et al.} classified six linguistic patterns from development emails~\cite{Di2016DECA}. We also employed this classification to classify linguistic patterns used to express architecture decisions in AI-based systems development (RQ1). For RQ2, RQ3, RQ4, and RQ5, we conducted a qualitative data analysis by applying the Constant Comparison method \cite{Glaser1965The}, with which each part of the data, e.g., emergent codes, categories, is constantly compared to all other parts of the data to explore differences and similarities in the data \cite{core2006lillemor}. Our data analysis process contains the following steps: (1) The first author and second author labelled the content of the filtered posts and issues with the corresponding data items in Table \ref{Data Items and RQ}; (2) The first author rechecked the labelling results of the second author to make sure that the extracted data were correctly labelled; (3) The first author combined all the codes into higher-level concepts and transformed them into categories. The third author then examined the analysis results and disagreements were eliminated through discussions with the first author. The data analysis methods used for answering the RQs are listed in Table~\ref{Data Items and the data analysis methods used for the research questions}. All the labelling and analysis results in the MS Excel sheets have been provided online~\cite{replpack}.

\begin{table}[htbp]
\caption{Data items and their analysis methods for answering the RQs}
\label{Data Items and the data analysis methods used for the research questions}
\begin{tabular}{m{0.25cm}<{\centering}m{3.15cm}m{3.3cm}m{0.4cm}<{\centering}}
\hline
\textbf{\#} & \textbf{Data Item}                         & \textbf{Data Analysis Method}                  & \textbf{RQ}  \\ \hline
D1          & Decision description                       & Predefined classification \cite{Di2016DECA}    & RQ1 \\ \hline
D2          & Decision type                              & Constant comparison                            & RQ2 \\ \hline
D3          & Application domain                         & Constant comparison                            & RQ3 \\ \hline
D4          & Quality attribute considered               & Constant comparison                            & RQ4 \\ \hline
D5          & Limitation and challenge                   & Constant comparison                            & RQ5 \\ \hline
\end{tabular}
\end{table}

\section{Results}
\label{sec:results}

In this section, we present the results of the five RQs formulated in Section~\ref{RQs}. We first illustrate how we analyzed the instances extracted from the 174 selected SO posts and 128 GitHub issues according to the data items listed in Table \ref{Data Items and RQ} in order to answer each RQ, and we then provide the final results of each RQ.\\

\noindent \textbf{RQ1: How do developers express architecture decisions made in AI-based systems development?}\\
To answer RQ1, we first collected 2,417 instances used by practitioners to express architecture decisions in AI-based systems development from SO and GitHub. We then used a predefined classification, the linguistic patterns proposed by Sorbo \textit{et al.}~\cite{Di2016DECA}, to classify the decision expressions according to their purposes. Finally, we extracted and classified architecture decisions expressed in natural language into six linguistic patterns. Table \ref{Linguistic patterns of architecture decisions in AI-based systems development with their examples, counts and percentages} shows the six linguistic patterns identified to express architecture decisions made in AI-based systems development with their examples and percentages.

More specifically, we found that \textit{Solution Proposal} and \textit{Information Giving} are the two most frequently used linguistic patterns, accounting for 31.8\% and 27.5\%, respectively. Whereas, \textit{Problem Discovery} and \textit{Opinion Asking} catch the least attention from the practitioners when describing decisions made in AI-based systems development.

\begin{table*}[htbp]
\caption{Linguistic patterns of architecture decisions in AI-based systems development}
\label{Linguistic patterns of architecture decisions in AI-based systems development with their examples, counts and percentages}
\begin{tabular}{m{2.4cm}m{12.8cm}m{0.6cm}<{\centering}m{0.6cm}<{\centering}}
\hline
\textbf{Linguistic Pattern}  & \textbf{Example}                                      & \textbf{Count}    & \textbf{\%} \\ \hline
Solution Proposal            & \textit{\texttt{The other solution} that we thought about was to send the output folder from the trainer to the ML Engine job, so that when our Module A asks for the status of that ML Engine Job, it gets the output folder.} (SO post \#45411311)   & 769  & 31.8\%  \\  \hline
Information Giving           & \textit{\texttt{I've} attempted to reduce the number of modules as I don't require all modules, \texttt{I have} also removed slices from the framework to just leave arm64.} (OpenCV issue \#13439)                                                    & 664  & 27.5\%  \\  \hline
Information Seeking          & \textit{\texttt{What is} the architecture of spaCy (modules/layers and so on)?} (spaCy issue \#5408)                                                                                                                                                   & 398  & 16.5\%  \\  \hline
Feature Request              & \textit{This is not a bug rather a \texttt{feature request}. As TensorFlow library provides TFLite models to run on Android, iOS platform, can we a build tfjs wrapper to allow tfjs to directly load TFlite model in browser. This will allow same model being used across multiple platforms.} (TensorFlow.js issue \#991)   & 251  & 10.4\%  \\  \hline
Problem Discovery            & \textit{\texttt{The main question} when using so many technologies is to scale well the infrastructure and the resources given to each one.} (SO post \#27730628)                                                                                      & 182  & 7.5\%   \\  \hline
Opinion Asking               & \textit{\texttt{Does} an object need to know how to update and draw itself, even having dependencies from higher levels that somehow are supplied to them?} (SO post \#5458760)                                                                        & 153  & 6.3\%   \\  \hline
\end{tabular}
\end{table*}

\begin{table*}[htbp]
\caption{Types of architecture decisions in AI-based systems development}
\label{Types of architecture decisions in AI-based systems development}
\begin{tabular}{m{3cm}m{12.2cm}m{0.6cm}<{\centering}m{0.6cm}<{\centering}}
\hline
\textbf{Type}                   & \textbf{Example}                                                                  &\textbf{Count}   & \textbf{\%}           \\ \hline
Technology Decision             & \textit{The general question is - what can be a resonable solution/technology/architecture of such system?} (SO post \#39253328)                                                                                                                                 & 61   & 21.3\%  \\ \hline
Component Decision              & \textit{What you are asking to do is a big project that consists of too many components and no one can really show you how to do everything like that.} (AirSim issue \#1744)                                                                                    & 59   & 20.6\%  \\ \hline
Data Decision                   & \textit{For video streaming, how should this c/s architecture program transmit data, and where should the decoder be placed?} (Triton issue \#4487)                                                                                                              & 38   & 13.3\%  \\ \hline
Service Decision                & \textit{I have a problem with this service design. I was thinking about creating the main service on Kubernetes that will pull messages from PubSub then this main service would create pods (or rather jobs) to run the actual ML work.} (SO post \#66225804)   & 21   & 7.3\%   \\ \hline
Implementation Decision         & \textit{Implement an efficient encoding of a knowledge base + all APIs / interfaces, to integrate with the current processing pipeline.} (spaCy issue \#3339)                                                                                                    & 19   & 6.6\%   \\ \hline
Deployment Decision             & \textit{We have deployed a non-managed Kubernetes cluster using an EC2 instance and we connect the Worker Nodes to the master using a VPN.} (SO post \#55886095)                                                                                                 & 18   & 6.3\%   \\ \hline
Tool Decision                   & \textit{Looking for any references architectures, or tooling that can be stiched together so we can incrementally improve our setup.} (SO post \#11587135)                                                                                                       & 18   & 6.3\%   \\ \hline
Concurrency Decision            & \textit{I'm developing a multiplayer game application with C++ and currently in the process of choosing an appropriate multithreading architecture for it.} (SO post \#1697820)                                                                                  & 17   & 5.9\%   \\ \hline
Integration Decision            & \textit{In our specific use case, we decided to build a flask server acting as a relay for incoming requests to Triton, meaning we have integrated the Triton Python HTTP client inside a Flask app.} (Triton issue \#2744)                                      & 12   & 4.2\%   \\ \hline
Architecture Pattern Decision   & \textit{I am also using a flask driven microservice architecture.} (AllenNLP issue \#1313)                                                                                                                                                                       & 11   & 3.8\%   \\ \hline
Design Pattern Decision         & \textit{While looking at this architecture, I am immediately thinking about a Chain of Responsibility Design Pattern.} (SO post \#63788496)                                                                                                                      & 5    & 1.7\%   \\ \hline
Containerization Decision       & \textit{When NM launches a container, move job container into k8s pod.} (OpenPAI issue \#2195)                                                                                                                                                                       & 4    & 1.4\%   \\ \hline
Decomposition Decision          & \textit{I'm guessing this will be addressed by a re-engineering of the demo architecture, separating different models?} (AllenNLP issue \#443)                                                                                                                   & 3    & 1.0\%   \\ \hline
\end{tabular}
\end{table*}

\vspace{0.2cm}

\noindent \textbf{RQ2: What types of architecture decisions are made in AI-based systems development?}\\
To answer RQ2, we manually identified the content related to architecture decisions in AI-based systems development from the collected posts and issues, and we then analyzed and summarized them in brief sentences. During the identification process, if the same decision was mentioned multiple times within a post or an issue, we counted it only once. Finally, we used the Constant Comparison method \cite{Glaser1965The} to form a category of architecture decisions made in AI-based systems development. Table \ref{Types of architecture decisions in AI-based systems development} presents the thirteen types of architecture decisions made in AI-based systems development with their counts and percentages.


As shown in Table \ref{Types of architecture decisions in AI-based systems development}, the top three types of decisions are \textit{Technology Decision}, \textit{Component Decision}, and \textit{Data Decision}, in which both \textit{Technology Decision} and \textit{Component Decision} account for over 20\%. The percentages of \textit{Service Decision}, \textit{Implementation Decision}, \textit{Deployment Decision}, \textit{Tool Decision}, \textit{Concurrency Decision}, \textit{Integration Decision}, and \textit{Architecture Design Pattern Decision} are between 7.3\% and 3.8\%. The percentages of \textit{Design Pattern Decision}, \textit{Containerization Decision}, and \textit{Decomposition Decision} are less than 2.0\%.

\vspace{0.2cm}

\noindent \textbf{RQ3: What are the domains of AI-based systems in which architecture decisions are made?}\\
To answer RQ3, we counted the application domains in which architecture decisions are made in AI-based systems development and got the results presented in Table \ref{The domains in which architecture decisions are made in AI-based systems development} with 145 instances, in which 106 instances are from SO posts and 39 instances are from GitHub issues.

We identified 18 application domains, in which \textit{Game} is dominant, accounting for more than 30\%. Next comes \textit{Data Processing}, which is more than 15\%. \textit{Image Processing} and \textit{Mobile Application} are both around 10\%. \textit{Unmanned System}, \textit{Real-time System}, \textit{Internet of Things (IoT)}, \textit{Audio Processing}, \textit{Biology and Chemistry}, \textit{Audio Processing}, \textit{E-Business}, \textit{Chatbot}, and \textit{Social Network} were presented between 2\textasciitilde9 times each (1.4\% to 6.2\%). The rest application domains, including \textit{Food industry}, \textit{Business Process Management}, \textit{Expert System}, \textit{Embedded System}, \textit{Mechatronics}, and \textit{Robotic System} were mentioned only once (0.7\%).

\vspace{0.2cm}

\noindent \textbf{RQ4: What quality attributes are considered when developers make architecture decisions in AI-based systems development?}\\
To answer RQ4, we extracted the data item D4 ``Quality attribute considered'' from the included SO posts and GitHub issues with 137 instances, and finally we categorized them into twelve types, which are presented in Table \ref{The quality attributes considered in AI-based systems development}.

\textit{Performance} is the most commonly considered quality attribute by practitioners, accounting for nearly 40\% of the total instances. Meanwhile, the proportions of \textit{Maintainability} and \textit{Ease of Implementation} are similar, both around 10\%. When making architecture decisions, practitioners also pay attention to \textit{Cost and Effort}, \textit{Scalability}, \textit{Flexibility}, \textit{Complexity}, \textit{Security}, \textit{Reliability}, and \textit{Usability}, which range from 4.4\% to 6.6\%. The percentages of both \textit{Robustness} and \textit{Persistence} are below 1.5\%, and they are the quality attributes least considered by practitioners.


\begin{table}[htbp]
\centering
\caption{The domains in which architecture decisions are made}
\label{The domains in which architecture decisions are made in AI-based systems development}
\begin{tabular}{m{2cm}m{0.6cm}<{\centering}m{0.6cm}<{\centering}|m{2cm}m{0.6cm}<{\centering}m{0.6cm}<{\centering}}
\hline
\textbf{Domain}                 & \textbf{Count}            & \textbf{\%}          & \textbf{Domain}                & \textbf{Count}          & \textbf{\%}           \\ \hline
Game                            & 45                        & 31.0\%               & E-Business                     & 3                       & 2.1\%                 \\ \hline
Data Processing                 & 24                        & 16.6\%               & Chatbot                        & 3                       & 2.1\%                 \\ \hline
Image Processing                & 17                        & 11.7\%               & Social Network                 & 2                       & 1.4\%                 \\ \hline
Mobile Application              & 14                        & 9.7\%                & Food Industry                  & 1                       & 0.7\%                 \\ \hline
Unmanned System                 & 9                         & 6.2\%                & Business Process Management    & 1                       & 0.7\%                 \\ \hline
Real-time System                & 7                         & 4.8\%                & Expert System                  & 1                       & 0.7\%                 \\ \hline
Internet of Things (IoT)        & 6                         & 4.1\%                & Embedded System                & 1                       & 0.7\%                 \\ \hline
Biology and Chemistry           & 5                         & 3.4\%                & Mechatronics                   & 1                       & 0.7\%                 \\ \hline
Audio Processing                & 4                         & 2.8\%                & Robotic System                 & 1                       & 0.7\%                 \\ \hline
\end{tabular}
\end{table}

\begin{table*}[htbp]
\caption{The quality attributes considered when making architecture decisions in AI-based systems development}
\label{The quality attributes considered in AI-based systems development}
\begin{tabular}{m{2.1cm}m{13.15cm}m{0.6cm}<{\centering}m{0.6cm}<{\centering}}
\hline
\textbf{Quality Attribute}   & \textbf{Example} & \textbf{Count} &\textbf{\%} \\ \hline
Performance            & \textit{However, a multi-threaded application would be more performant, as it would allow parallel usage without incurring a large memory requirement per "thread".} (AllenNLP issue \#443)                           & 51  & 37.2\% \\ \hline
Maintainability        & \textit{It would hopefully cause it to be more maintainable AND have the added side effect of implementing your scripting interface for you.} (SO post \#609076)                                                      & 19  & 13.9\% \\ \hline
Ease of Implementation & \textit{Why can't OpenCV.org develop a simple module of OpenPose so that it can be easily imported, and effectively call some functions like find Hands location or draw the skeleton etc.} (OpenCV issue \#17271)    & 13  & 9.5\%  \\ \hline
Cost and Effort        & \textit{Deploying Sagemaker to the same account that holds the rest of the business infrastructure is critical to avoid infrastructure overhead.} (MLflow issue \#4931)                                               & 9   & 6.6\%  \\ \hline
Scalability            & \textit{AWS is a great fit for the application you describe. AWS Fargate/RDS will host your Django application. You have the option of using AWS Batch to handle your processing. One huge advantage is the ability to scale according to the needs of your application.} (SO post \#64390919) & 8   & 5.8\%  \\ \hline
Flexibility            & \textit{TensorLayer provides the access to the native APIs of TensorFlow and therefore help users to achieve a flexible control within the engine.} (TensorLayer issue \#185)                                         & 8   & 5.8\%  \\ \hline
Complexity             & \textit{It all depends on the complexity your are ready to put in your project.} (SO post \#64892374)  & 8   & 5.8\%  \\ \hline
Security               & \textit{If your model is non-generic and you want to protect it from being copied, this also has added security benefits since users won't have access to the model itself.} (SO post \#55065392)                     & 6   & 4.4\%  \\ \hline
Reliability            & \textit{Generally, people opt for a message bus like Kafka to handle back-pressure from slow downstream consumers and also to provide reliability (by persisting to disk) to prevent data loss.} (SO post \#39253328) & 6   & 4.4\%  \\ \hline
Usability              & \textit{We will learn more about the usability and improve in the future.} (Kubeflow issue \#5509)    & 6   & 4.4\%  \\ \hline
Robustness             & \textit{It also has a very compelling convention over configuration mechanism (...). This is more powerful than any other mechanism I've seen.} (SO post \#10799101)                & 2   & 1.5\%  \\ \hline
Persistence            & \textit{However even if location entities are nicely isolated there another problem - persistence.} (SO post \#1697820)  & 1   & 0.7\%  \\ \hline
\end{tabular}
\end{table*}

\begin{table*}[htbp]
\caption{Limitations and challenges of making architecture decisions in AI-based systems development}
\label{Limitations and challenges of making architecture decisions in AI-based systems}

\begin{tabular}{m{1.75cm}m{13.5cm}m{0.6cm}<{\centering}m{0.6cm}<{\centering}}
\hline
\textbf{Limitation \& Challenge}   & \textbf{Example}        & \textbf{Count}   & \textbf{\%} \\ \hline
Design Issue                       & \textit{You made a valiant effect in the past which unfortunately stranded on architectural design. I don't see the use case for plotting the mean of several channels.} (MNE-Python issue \#3393)                                                                                           & 22      & 59.2\%   \\ \hline
Data Issue                         & \textit{Then how should I architect my application in such a way that it should take care of scale-ability in term of supporting thousands of user logging huge amount of data every mill second.} (SO post \#52494424)  & 16     & 32.7\%    \\ \hline
Integration Issue                  & \textit{Unfortunately, it was not clear how to integrate DL4J within what we currently have or if it simply replicates the stack I have now.} (SO post \#47803989)                                                       & 1      & 2.0\%     \\ \hline
Programming Language Issue         & \textit{I could not find it spelled out in the documentation: Do AI Platform custom containers only support Python ML frameworks, or are other languages also supported?} (SO post \#61041666)                           & 1      & 2.0\%     \\ \hline
Protocol Issue                     & \textit{Perhaps the AIs run as separate processes, and communicate with a central "board" process via sockets? I don't think there is a common protocol.} (SO post \#37369415)                                           & 1      & 2.0\%     \\ \hline
Deployment Issue                  & \textit{If I want to deploy my project including launch of multiple deep models built with different frameworks (Pytorch, tensorflow) then what's good option for that?} (SO post \#65785528)                                                                & 1      & 2.0\%     \\ \hline
\end{tabular}
\end{table*}

\vspace{0.2cm}

\noindent \textbf{RQ5: What are the limitations and challenges of making architecture decisions in AI-based systems development?}\\
To answer RQ5, a total of 49 instances were collected and analyzed to derive seven types of limitations and challenges encountered when practitioners made architecture decisions in AI-based systems development, as presented in Table \ref{Limitations and challenges of making architecture decisions in AI-based systems}.


During architecture decision-making in AI-based systems development, the major limitations and challenges are \textit{Design Issues} related to AI-based systems and \textit{Data Issues} related to data storage \& processing, which together account for more than 90\%. The rest limitations and challenges, including \textit{Integration Issue}, \textit{Programming Language Issue}, \textit{Protocol Issue}, and \textit{Deployment Issue}, were only mentioned once.

\section{Discussion}
\label{sec:discussion}

\subsection{Interpretation of Results}
\label{subA}
\noindent\textbf{RQ1: How do developers express architecture decisions made in AI-based systems development?}

As shown in Table \ref{Linguistic patterns of architecture decisions in AI-based systems development with their examples, counts and percentages}, the natural language expressions of the identified architecture decisions made in AI-based systems development can be classified into six predefined linguistic patterns according to the purposes of decision expressions \cite{Di2016DECA}, in which \textit{Solution Proposal} and \textit{Information Giving} are most frequently used. \textit{Solution Proposal} suggests possible solutions as architecture decisions and \textit{Information Giving} is employed to provide others with information related to decision-making. Owing to the wide variety of AI technologies, for one specific architectural problem, there may be many potential solutions, and practitioners tend to use \textit{Solution Proposal} to provide possible solutions. As developers/architects need to provide relevant information to clarify their problems related to AI-based systems development when making architecture decisions, \textit{Information Giving} can be used to inform or update others about the required information for making decisions.

\noindent\textbf{RQ2: What types of architecture decisions are made in AI-based systems development?} 

As shown in Table \ref{Types of architecture decisions in AI-based systems development}, there are thirteen types of architecture decisions identified in AI-based systems development. One of the most common types of architecture decisions is \textit{Technology Decision}, indicating that developers often need to decide which AI technologies to choose when developing AI-based systems. Another type of architecture decisions frequently made by practitioners is \textit{Component Decision}. One reason could be that \textit{Component Decisions} are frequently made in all types of systems and developers need to always consider how to design the system in, e.g., modules and layers. Besides, developers need to well design the components of AI-based systems so that AI-related components can work efficiently with non-AI related components of the systems. \textit{Data Decisions} are the third most frequently made decisions, showing that practitioners care more about data storage and processing when developing AI-based systems due to large and constantly changing data brought by training AI models.

\noindent\textbf{RQ3: What are the domains of AI-based systems in which architecture decisions are made?} 

As shown in Table \ref{The domains in which architecture decisions are made in AI-based systems development}, architecture decisions in AI-based systems development cover a total of eighteen domains, in which \textit{Game} is dominant. Many game software requires the use of AI-related components (e.g., AI engines), and AI-based systems are in great demand in game industry. It is reasonable that \textit{Game} is the major domain of AI-based systems in which architecture decisions are made. When practitioners make decisions in game systems with AI technologies, they make \textit{Component Decisions} most frequently. This is because the pervasive use of AI-related components (e.g., AI characters) in game systems leads to intensive interactions between AI-related and non-AI related components, which consequently raises more design issues related to components. Note that, there is no project from the \textit{Game} domain in the GitHub projects we selected, as the top 32 projects we selected based on the number of issues have no \textit{Game} applications. 

\noindent\textbf{RQ4: What quality attributes are considered when developers make architecture decisions in AI-based systems development?} 

As shown in Table \ref{The quality attributes considered in AI-based systems development}, we captured the quality attributes considered by practitioners when they make architecture decisions in AI-based systems development and mapped the considered quality attributes into twelve types. \textit{Performance}, accounting for more than 35\%, is the dominant quality attribute considered by practitioners. \textit{Performance} is related to time behaviour, resource utilization, and capacity of a system, which highly affects how well a system operates after implementation \cite{international2011systems}. According to our study results, for the quality attribute \textit{Performance}, practitioners are most concerned about the training and computing time of machine learning and deep learning models, and the speed of data processing in architecture decision-making of AI-based systems development, as one developer proposed ``\textit{a suggestion would be to mix and match encoder and decoder arch. (...) I'm still running a few experiments to figure out the best way to implement a RNN decoder (prioritizing performance to speed tradeoff)}'' (Fairseq issue \#2237).

\noindent\textbf{RQ5: What are the limitations and challenges of making architecture decisions in AI-based systems development?} 

As shown in Table \ref{Limitations and challenges of making architecture decisions in AI-based systems}, practitioners encounter various types of limitations and challenges during architecture decision-making of AI-based systems development, in which \textit{Design Issues} and \textit{Data Issues} are the main ones. Due to the introduction of many AI-related components, the system design of AI-based systems becomes challenging for developers, and they need to make \textit{Component Decisions} related to system structure design frequently, as mentioned in a GitHub issue ``\textit{what you are asking to do is a big project that consists of too many components and no one can really show you how to do everything like that}'' (AirSim issue \#1744). Developers want to develop an AI-based system that works efficiently and with a simple architecture. To achieve this, they need to design the AI-based system by following good design principles, which should be flexible and cannot be over-designed, as one practitioner mentioned that ``\textit{make your design flexible enough without making it overly complicated}'' (SO post \#68461674). \textit{Design Issues} include both AI specific limitations \& challenges (e.g., integrating AI-related components and deploying ML/DL models) and traditional architecture limitations \& challenges (e.g., component coupling), which is in line with the finding in \cite{serban2021empirical} that ``\textit{traditional architecture challenges play an important role when using ML components}''. \textit{Data Issues} are mainly caused by the large amount of data in AI-based systems due to the use of ML/DL models. Many developers had difficulties in choosing a better solution related to data processing and storage among a few approaches due to their unfamiliarity with data-related technologies, as one developer asked ``\textit{how to pick up the right technologies}'' because ``\textit{it's difficult to understand the pros and cons of each of them}'' (SO post \#45739159).

\subsection{Implications}

\noindent\textbf{\textit{Problem Discovery} and \textit{Opinion Asking} can be considered by practitioners in describing their architecture decisions in AI-based systems development}: 
In AI-based systems development, developers often encounter situations when the solutions that work for common problems may not be applicable to specific issues due to the introduction of AI-related components, as mentioned in one post ``\textit{my problem is the `big picture' architecture of this system: I see that all the pieces already exist (cloudera+mahout) but I'm missing a simple integrated solution for all my needs}'' (SO post \#13747670). In this context, developers need to specify architecture problems or unexpected behaviours using \textit{Problem Discovery}. For certain architecture issues, the choices of AI technologies are extensive, as one developer asked ``\textit{ML train/test jobs with data as huge S3 csv or directly from a preprocessed Relational or NoSql DB}'' (SO post \#68913355). Developers can then inquire of others to clearly show his/her point of view about possible decisions using \textit{Opinion Asking}. These two linguistic patterns that can assist practitioners in gaining others' help in their decision-making are seldom used, and we encourage practitioners to use these two linguistic patterns properly in describing their architecture decisions in AI-based systems development when they need informed suggestions from others about the decisions made.
 
\noindent\textbf{Making \textit{Technology Decisions} and \textit{Component Decisions} becomes challenging for practitioners in AI-based systems due to the introduction of AI-related components}: \textit{Technology Decision} is one of the major types of architecture decisions in AI-based systems development, which highly affects the technical characteristics of AI-based systems. Many practitioners have difficulties in choosing suitable technologies for AI-based systems due to the diversity and fast-changing speed of AI technologies, for example, they need to make a decision between Spark and Flink as a framework for an AI-based system (SO post \#44079728). This finding has been confirmed by Franch \textit{et al.}~\cite{franch2022architectural} that ``New Technology'' as an architecture decision type identified after analyzing 41 studies on architecture decisions in AI-based systems. \textit{Component Decision} is another major architecture decision type, and it is expected to provide guidelines for the interface and interaction design between AI-related and non-AI related components. 

\noindent\textbf{Practitioners should pay more attention to component design when making architecture decisions in developing AI-based \textit{Games}}: 
\textit{Game} is the most popular application domain of AI-based systems in which architecture decisions are made, requiring practitioners to pay more attention to making \textit{Component Decisions} in order to appropriately establish the communication between AI-related and non-AI related components of the system structure (see the interpretation of RQ3 results in Section \ref{subA}). Therefore, more research on component design in AI-based games can be beneficial for practitioners in the game industry.

\noindent\textbf{Researchers can explore how the quality attributes are prioritized and considered when making architecture decisions in AI-based systems development}: 
In total, we identified twelve types of quality attributes to consider when making architecture decisions in AI-based systems development, and practitioners need to make trade-offs between them. Though our work has identified the quality attributes that affect the architecture decision-making of AI-based systems development, we did not dive deeper into how practitioners prioritize these quality attributes and how these quality attributes impact architecture decision-making, which highly determines how the final system will be implemented and how well it operates.


\noindent\textbf{Limitations and challenges when making architecture decisions in AI-based systems development should be properly confronted}: When making architecture decisions in AI-based systems development, encountered limitations and challenges are highly specific to the characteristics of AI-based systems. The biggest limitations and challenges for practitioners are \textit{Design Issues}. Practitioners need a high learning curve to study diverse AI technologies in order to choose the suitable one for the system under development. Moreover, training ML/DL models makes data processing to be challenging in AI-based systems compared with traditional software systems, leading to a dominance of issues related to data. These limitations and challenges are a double-edged sword. If practitioners can treat them correctly, these restrictions and possible challenges can bring new opportunities for making architecture decisions in AI-based systems development, resulting in a successful software system, as one developer stated that ``\textit{make your design much more flexible and save you massive headaches later}'' (SO post \#1989152). On the contrary, if practitioners fail to deal with these limitations and challenges in a decent way, the difficulty of developing AI-based systems would increase.

\section{Threats to Validity}
\label{sec:threats}


\textbf{Construct validity} is the extent to which a measure can account for the theoretical structure and characteristics of the measure. There are three threats to the construct validity of this study: (1) The search terms used to collect related posts from SO and related issues from GitHub. This is a keyword-based search, and one possible threat to construct validity is that the results got from the search terms we used may not cover all the relevant posts and issues. Through applying the systematic approach proposed by Bosu \textit{et al.}~\cite{Bosu2014Identifying} to find synonyms and adding the abbreviations of machine learning and deep learning, we have partially reduced the degree of this threat. (2) Manual labelling, extraction, and analysis of data. These three steps are manually conducted, which may induce personal bias. To reduce the threat of data labelling, we conducted separate pilot data labelling of SO posts and GitHub issues before the formal data labelling, and we formulated the inclusion and exclusion criteria about whether a candidate post or issue should be included or not. To alleviate the threat of data extraction, the first author rechecked the data extraction results after the data extraction was conducted jointly with the second author, and any disagreements were discussed and resolved with the third author. To mitigate the threat of data analysis, the first author continuously consulted with the third author during the analysis process to reach an agreement. 

\textbf{External validity} notes the degree of generalization of the study results, which indicates whether the findings are representative and can be verified in similar contexts. The main threat to the external validity is the selection of data sources. To mitigate this threat, we chose two popular developer communities, SO and GitHub, as our data sources to extract architecture decisions in AI-based systems development. SO and GitHub have been widely used in empirical software engineering studies (e.g., \cite{islam2020repairing}, \cite{chen2021empirical}). Previous studies \cite{humbatova2020taxonomy}, \cite{aghajani2022software} found that the findings from SO and GitHub can be well validated by practitioners, which makes us believe that choosing these two data sources can be representative about architecture decisions in AI-based systems development.

\textbf{Reliability} refers to how well a particular research method can produce consistent results. To reduce this threat, before the formal data labelling, two authors conducted a pilot data labelling on 60 SO posts and 10 GitHub issues independently. We acknowledge that due to the small number of posts and issues this threat might still exist.
The Cohen's Kappa coefficient of the pilot labelling was 0.727 for SO posts and 0.737 for GitHub issues, both higher than 0.7, which indicates a decent agreement between the authors. In the process of data labelling, data extraction, and data analysis, any inconsistencies were resolved through discussion between the first author and other researchers. In addition, we provided the study dataset containing all the extracted data and labelling results of SO posts and GitHub issues for validation \cite{replpack}. 

\section{Related Work}
\label{sec:relatedWork}

\subsection{AI-based Systems Development}
Several studies have focused on the topic of AI-based systems development. Washizaki \textit{et al.}~\cite{washizaki2019studying} conducted a systematic literature review, which categorised and detailed a collection of software (anti-)patterns for machine learning systems. Lwakatare \textit{et al.}~\cite{lwakatare2019taxonomy} carried out an empirical investigation into software engineering challenges for machine learning systems, getting a taxonomy of common issues, including data dependency management, deployment difficulties and result reproduction challenges, without formally modelling design decisions. Similarly, Warnett and Zdun~\cite{warnett2022architectural} conducted a qualitative investigation into the technical challenges faced by practitioners in machine learning system deployment. They modelled current practices in machine learning deployment, leading to a UML-based architectural design decision model and identified architectural design decisions, various relations among them, decision options, and decision drivers in different sources. Gu \textit{et al.}~\cite{Gu2021Demystifying} extracted and analyzed real-world developers’ issues in distributed training of deep learning systems from SO and GitHub. They constructed a fine-grained taxonomy consisting of 30 categories regarding the fault symptoms and summarized common fix patterns for different symptoms. Their research tends to focus on a particular area of AI technologies, for example, machine learning or deep learning, and the studied systems are specialized AI systems. However, our study focuses on AI-based systems development, i.e., the systems that employ AI-related components, including the architecture decision expressions, decision types, involved application domains, considered quality attributes, and encountered limitations and challenges in AI-based systems development.

\subsection{Decisions in Software Engineering}
Decisions are either explicitly presented in knowledge management tools or implied in various textual artifacts. Many researchers have analyzed the decision-making in the process of software development. Hesse \textit{et al.}~\cite{2016Documented} studied the decision-making strategies and knowledge documented in issue reports of open source software, and the relationship between these strategies and knowledge. Li \textit{et al.}~\cite{Li2019Decisions} analyzed the posts from the Hibernate developer mailing list and manually extracted decisions to explore decision expressions, types of decisions, rationale behind decision-making, related artifacts, approaches used in decision-making, and the trend of decision-making in a time perspective. Olsson \textit{et al.}~\cite{olsson2019empirical} combined both quantitative and qualitative data to analyze the decision-making for quality requirements in a large multi-national company. 

\subsection{Architecture Decisions in AI-based Systems Development}
Franch \textit{et al.}~\cite{franch2022architectural} conducted a literature study on architecture decisions in AI-based systems by analyzing 41 related papers, and proposed a preliminary ontology for architectural decision-making. They presented a summary of concepts related to architecture decisions, system contexts, impacts, architectural elements, AI-related architectural elements, and architectural views, with decision types and quality attributes related to our RQ2 and RQ4. They emphasized that further investigation of the current state of research and practice in this area is needed, which confirms the motivation of our study. Serban and Visser \cite{serban2021empirical} conducted a mixed-methods empirical study with a systematic literature review, semi-structured interviews, and a qualitative survey to investigate the challenges and solutions for (re-)architecting systems with ML components. They also studied the quality attributes associated with decision drivers, while we conducted an empirical study by mining software repositories to identify the quality attributes that practitioners would consider when making architecture decisions in AI-based systems development.


Different from the work above, we studied the expressions and types of architecture decisions, application domains, quality attributes considered, and limitations and challenges faced when making architecture decisions in AI-based systems development through mining the data from SO and GitHub.





\section{Conclusions}
\label{sec:conclusions}

AI-related components are becoming a fundamental part of current software systems, and AI-based systems have been extensively used in our daily live. Architecture and architecture decisions play an important role in software systems. We conducted an empirical study on Stack Overflow (SO) and GitHub to explore architecture decisions in AI-based systems development from practitioners' perspectives. We used a keyword-based search to collect data from SO and GitHub, and finally got 174 posts and 128 GitHub issues, and we manually extracted architecture decisions related to AI-based systems to explore decision expressions, decision types, involved application domains, quality attributes considered, and limitations and challenges encountered in architecture decision-making of AI-based systems development. The main results are that: (1) All the architecture decisions are expressed in six linguistic patterns, and the leading patterns are \textit{Solution Proposal} and \textit{Information Giving}. (2) The main types of architecture decisions are \textit{Technology Decision}, \textit{Component Decision}, and \textit{Data Decision}. (3) The major domain of AI-based systems in which architecture decisions are made is \textit{Game}. (4) The most common quality attribute considered when making architecture decisions in AI-based systems development is \textit{Performance}. (5) \textit{Design Issues} and \textit{Data Issues} are the most frequently encountered limitations and challenges when making architecture decisions in AI-based systems development. Our future work aims to address the identified limitations and challenges when making architecture decisions in AI-based systems development.



\balance

\bibliographystyle{ieeetr}
\bibliography{references}

\end{document}